\documentclass[pre, reprint]{revtex4-1}
\usepackage{amsmath}
\usepackage{amssymb}
\usepackage[ansinew]{inputenc}
\usepackage[english]{babel}
\usepackage{graphicx}
\usepackage{color}
\begin{document}
\newcommand{\symmS}{\mathcal{S}}
\newcommand{\ndc}{\textsc{ndc}}
\newcommand{\anc}{\textsc{anc}}
\newcommand{\ssl}{\textsc{ssl}}
\newcommand{\D}{\mathrm{d}}
\newcommand{\E}{\mathrm{e}}
\newcommand{\I}{i}
\newcommand{\order}{\mathcal{O}}
\newcommand{\td}[2]{\frac{\D #1}{\D #2}}
\newcommand{\dt}[1]{\td{#1}{t}}
\newcommand{\pd}[2]{\frac{\partial #1}{\partial #2}}
\newcommand{\pdin}[2]{\partial #1 / \partial #2}
\newcommand{\usc}{{u_\text{sc}}}
\newcommand{\uext}{{u_\text{ext}}}
\newcommand{\duext}{{{\dot u}_\text{ext}}}
\newcommand{\dusc}{{{\dot u}_\text{sc}}}
\newcommand{\wsdc}{{\omega_\text{sdc}}}
\newcommand{\ws}{{\omega_\text{s}}} 
\newcommand{\wpl}{\omega_\text{pl}}
\newcommand{\wplp}{{\tilde \omega}_\text{pl}}
\newcommand{\eps}{\varepsilon}
\newcommand{\weq}{w_\text{eq}}
\newcommand{\gw}{\gamma_w}
\newcommand{\gv}{\gamma_v}
\newcommand{\dc}{\textsc{dc}}
\newcommand{\ac}{\textsc{ac}}
\newcommand{\Eext}{{E_\text{ext}}}
\newcommand{\Esc}{{E_\text{sc}}} 
\newcommand{\Weq}{{W_\text{eq}}}
\newcommand{\OmegaR}{\Omega_\text{R}}
\newcommand{\OmegaS}{\Omega_\text{S}}
\newcommand{\OmegaH}{\Omega_\text{H}}
\newcommand{\OmegaM}{\Omega_\text{M}}
\newcommand{\OmegaHM}{\Omega_{\text{H}, \text{M}}}
\newcommand{\diag}{\mathop{\mathrm{diag}}}
\newcommand{\changed}[1]{{\color{blue}#1}}
\newcommand{\cang}{\phi^*}
\newcommand{\cvar}{\mathcal{V}}
\newcommand{\pz}{p_z}
\newcommand{\ie}{\textit{i.e.}}
\newcommand{\eg}{\textit{e.g.}}
\title{Devil's staircase, spontaneous-DC bias, and chaos via
  quasiperiodic plasma oscillations in semiconductor superlattices}
\date{\today}
\author{Jukka Isohätälä}
\affiliation{Department of Physical Sciences, P.O. Box 3000, FI-90014 University of Oulu, Finland}
\author{Kirill N. Alekseev}
\affiliation{Department of Physical Sciences, P.O. Box 3000, FI-90014 University of Oulu, Finland}
\begin{abstract}
  We study a plasma instability in semiconductor superlattices
  irradiated by a monochromatic, pure \ac\ electric field.  The
  instability leads to sustained oscillations at a frequency
  $\omega_2$ that is either incommensurate to the drive, or
  frequency-locked to it, $\omega_2 = (p/q) \cdot \omega$.  A
  spontaneously generated \dc\ bias is found when either $p$ or $q$ in
  the locking ratio are even integers. Frequency locked regions form
  Arnol'd tongues in parameter space and the ratio $\omega_2 / \omega$
  exhibits a Devil's staircase.  A transition to chaotic motion is
  observed as resonances overlap.
\end{abstract}
\maketitle
\section{Introduction} 
Miniband transport in semiconductor superlattices (\ssl) is a fertile
ground for observing various nonlinear transport
phenomena\cite{wacker02:ssreview}.  Miniband \ssl s subject to
intense, monochromatic THz-radiation exhibits a variety of such
effects including dissipative chaos \cite{alekseev96:dissp-chaos-ssl,
  cao00:chaos-qdot-miniband, romanov00}, and generation of a quantized
spontaneous \dc\ bias \cite{alekseev98:spont-dc}. A sufficient,
although not necessary, requirement for the appearance of such novel
dynamics is the presence of instabilities in simpler types of
motion. A well-studied example of this is the instability occurring
near the conditions of negative differential conductance (\ndc). This
is known to lead to the generation of a nearly quantized spontaneous
\dc\ bias via dynamical breaking of symmetry\cite{alekseev98:spont-dc,
  alekseev02a, alekseev05:lsslcr, isohatala05:sbpend, isohatala10},
but also domains of different electric field strength that invalidate
the used models\cite{ktitorov, *ktitorov-trans, buttiker77}. The
conversion of pure THz \ac\ input to \dc\ output, or rectification,
has immediate applications to detection of the radiation, and
therefore finding instabilities that might lead to such broken
symmetry is of great interest.

In Ref.~\onlinecite{romanov79, *romanov79-trans} an instability
different from that observed at \ndc\ was reported. A weak probe
\ac\ field with a frequency that is an irrational multiple of the pump
frequency was found to be divergent provided the degree of
nonlinearity was sufficiently high. Notably this instability occurs
outside the regions of \ndc\ implying the absence of domains. In this
paper we report our findings on the same phenomenon but using a more
elaborate model that allows us to account for all harmonics in the
presence of nonlinearity, study the problem in terms of the applied
field amplitude $E_0$ and frequency $\omega$, and importantly consider
the complex dynamics that arise following this instability. Our goal
is twofold, firstly to understand the dynamics of miniband electrons
with strong nonlinearity and secondly, apply the instability for the
detection of THz radiation via, for instance, rectification.

As our model, we will use the sinusoidal miniband superlattice balance
equations \cite{ignatov95, alekseev96:dissp-chaos-ssl,
  alekseev98:spont-dc, alekseev98:symmbr} with the self-consistent
electric field:
\begin{subequations}
  \label{eq:be}
  \begin{eqnarray}
    \dot V &=& -\frac{e a^2}{\hbar^2} E W 
               - \frac{1}{\tau} V,  \label{eq:be-v} \\
    \dot W &=& e E V - \frac{1}{\tau}(W - \Weq),\label{eq:be-w} \\
    {\dot E}_\text{sc} &=& -\frac{4\pi e N}{\eps_0} V, \label{eq:be-Esc}.
  \end{eqnarray}
\end{subequations} 
The variables $V$ and $W$ are the average velocity and kinetic energy
along the superlattice axis of the electron ensemble following a
distribution given by the Boltzmann transport equation.  $E$ is the
total electric field inside the superlattice and includes the
externally applied sinusoidal and the self-consistent fields,
\begin{equation}
  E(t) = E_0 \cos \omega t + \Esc(t).
  \label{eq:e}
\end{equation}
The first two equations of Eqs.~(\ref{eq:be}) are the well-known
superlattice balance equations for a sinusoidal
miniband\cite{wacker02:ssreview, ignatov95} with a miniband width of
$\Delta_0$, superlattice period $a$, and equilibrium average energy
$\Weq$, and with dissipation modeled using constant phenomenological
relaxation rate $\gamma = 1/\tau$.  Constants $e$ and $\eps_0$ are the
electron charge and relative permittivity, respectively. The model
includes the effects of displacement
currents\cite{alekseev96:dissp-chaos-ssl, romanov01} via
Eqs.~(\ref{eq:e}) and~(\ref{eq:be-Esc}).  $\Esc$ connects the total
electric field to the electric current, and thus introduces an
additional degree of nonlinearity to the problem. The strength of the
current to $E$-field coupling is proportional to the electron density
$N$ and the maximum velocity allowed by the superlattice miniband,
$V_\text{max} = a \Delta_0 / 2\hbar$.  This is conveniently expressed
by a single parameter, the plasma frequency $\wpl$:
\begin{equation}
  \wpl^2 = \frac{2\pi e^2 N a^2 \Delta_0}{\hbar^2 \eps_0}.
\end{equation}
The product $\wpl \tau$ then controls the balance of dissipation and
nonlinearity.

\newcommand{\poinc}{G}

\begin{figure}
  \begin{center}
    \includegraphics[scale=0.95]{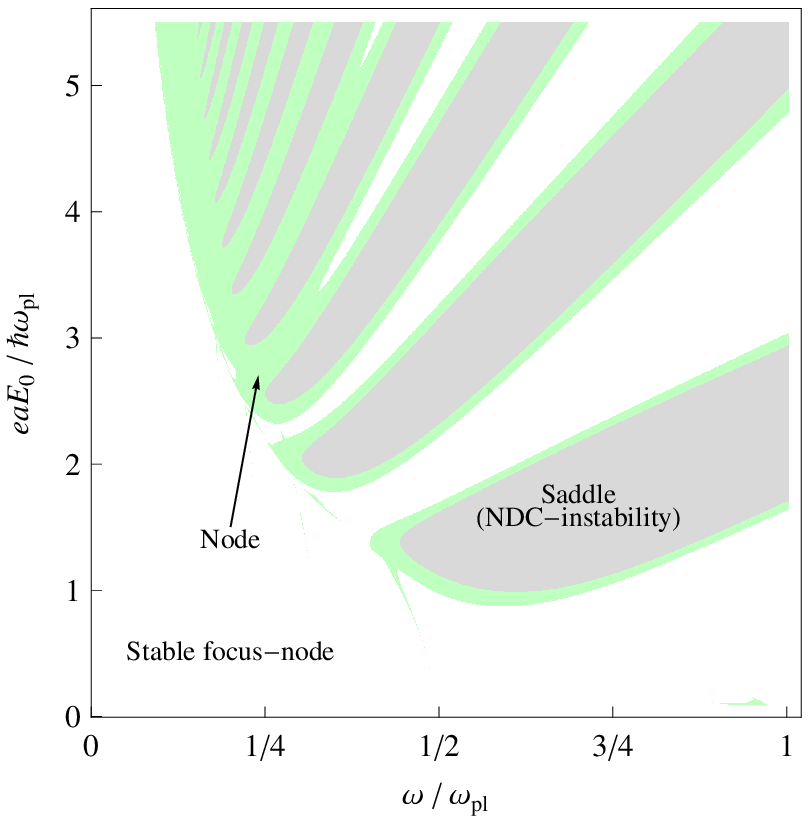}
    \includegraphics[scale=0.95]{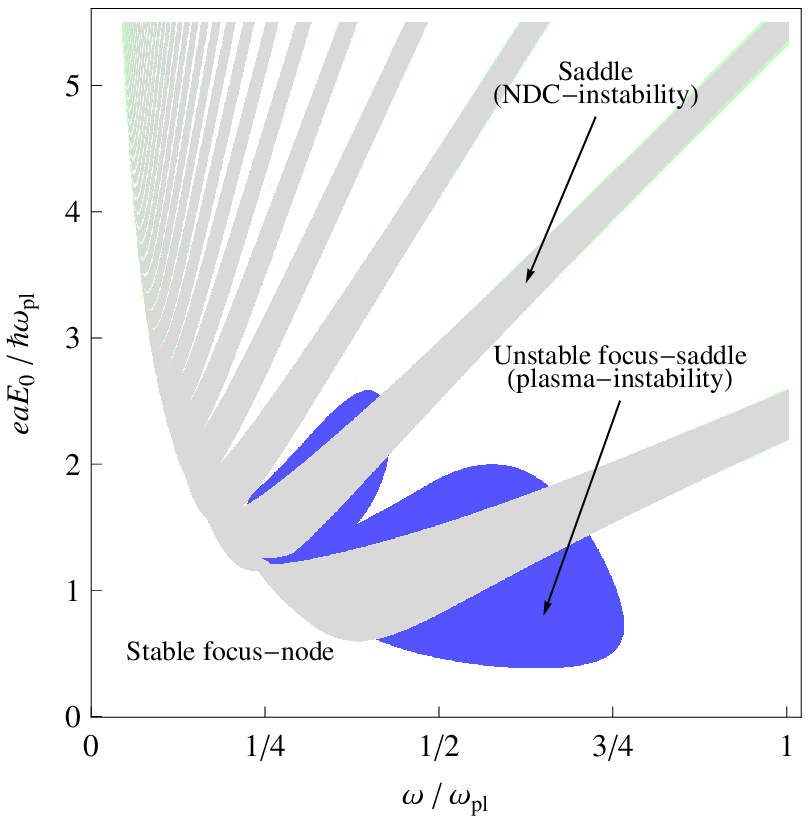}
  \end{center}
  \caption{\label{fig:fpr} [Color online] Types of fixed-points of the
    map $\poinc$, (a) $\wpl \tau = 3$, (b) $\wpl \tau = 12$.  For all
    $\wpl \tau$ the \ndc-instability is present, visible here as the
    regions where the fixed-point is a saddle [gray]. Equilibrium is a
    stable focus-node [white] for the rest of the shown parameter
    range, except near the borders of \ndc-instability, where it is a
    node [green]. In (a) the \ndc\ type is the only instability, but
    in (b) also the focus can destabilize, resulting in the region
    where the point is an unstable focus-saddle [blue].}
\end{figure}

\section{Plasma instability}
Our focus are the instabilities occurring in the symmetric limit-cycles
of Eqs.~(\ref{eq:be}). By symmetry we here refer to the invariance of
the equations under the transformation $\symmS$:
\begin{equation}
  \symmS: (t, V, W, \Esc) \to (t + T/2, -V, W, -\Esc).
  \label{eq:symm}
\end{equation}
A symmetric limit-cycle is then to be understood as a cycle that is
invariant in $\symmS$: $(V(t+T/2), W(t+T/2), \Esc(t+T/2)) = (-V(t),
W(t), -\Esc(t))$.  Symmetric cycles represent a basic type of dynamics
that do not support effects such as spontaneous generation of
\dc\ bias. Our interest in them lies in the fact that a loss of
stability in such a cycle would imply a transition to some different
types of dynamics, provided no other symmetric cycles are stable for
the same parameters. The stability of symmetric cycles can be studied
by considering the fixed-points of the map $\poinc$: $\poinc(X(t)) =
SX(t + T/2)$, where $X(t) = (V(t), W(t), \Esc(t))$, and $S =
\diag(-1,1,-1)$. Clearly, a fixed point of $G$ corresponds to a
symmetric cycle of Eqs.~(\ref{eq:be}). To map the parameter-space
regions where a symmetric limit-cycle loses its stability, we have
computed the fixed-points (equilibria) of $\poinc$ which we then
characterize according the eigenvalues of the Jacobian $J_{ij} =
\partial \poinc_i / \partial X_j$. 

Our findings are shown in Fig.~\ref{fig:fpr} where we have plotted the
type of the equilibrium on $(\omega, E_0)$-parameter plane for two
different degrees of nonlinearity, $\wpl \tau = 3$ and $\wpl \tau =
12$. We have also set $\Weq = -\Delta_0/2$. Two types of stable
equilibria are present for both cases. Typically, the equilibrium is a
focus-node, implying that the steady state is reached via damped
oscillations. Additionally, also node-type equilibria exist. For
these, trajectories follow exponentially converging, non-oscillatory
orbits onto the limiting motion.

Turning to the unstable fixed points, for plasma frequencies $\wpl
\tau \lesssim 8$ the only unstable fixed-point is a saddle.  These
appear as the exponential convergence to a node changes to exponential
divergence, and can be shown to coincide with \ndc. This is shown
explicitly in App.~\ref{app:math}. Since here the \ndc\ appears at
zero \dc\ voltage, a region of absolute negative conductivity (\anc)
is always associated with its appearance. This leads to dynamical
rectification\cite{dunlap93:bo-f2v-conv, ignatov95}, an effect that is
studied in detail elsewhere\cite{alekseev98:spont-dc, romanov00,
  romanov01, isohataladc}. We refer to the saddle-type instability as
the \ndc-instability. This instability persists for all $\wpl \tau$,
and occurs approximately for $(\omega, E_0)$ such that $eaE_0 / (\hbar
\omega)$ is a root of Bessel $J_0$ function\cite{ignatov95}.

As the plasma frequency is increased also the stable focus-node loses
its stability, turning into an unstable focus-node in a
Hopf-bifurcation.  At the Hopf-bifurcation point the damped
oscillations around a focus-node become diverging.  This can be
identified as the plasma instability of Ref.~\onlinecite{romanov79,
  *romanov79-trans} as it describes just such dynamics.  Importantly,
the it indeed appears outside the regions of \ndc, therefore
suggesting that at least electric domains will not form. The lowest
degree of nonlinearity that is required is $\wpl \tau \sim 8$,
although values in the excess of $\sim 10$ are required to observe the
effect in a reasonably large range of parameters.
We note that the signature of plasma oscillations is visible for a
much wider range of $\wpl\tau$ as the damped oscillations
corresponding to a focus-node -type equilibrium.

In terms of superlattice parameters, the requirement of high $\wpl
\tau$ can be achieved for instance for a superlattice with $n \simeq
10^{18}\;\text{cm}^{-3}$ and $\Delta_0 \simeq
100\;\text{meV}$, with other parameters being $a = 6\;\text{nm}$,
$\gamma \simeq 4\;\text{THz}$ ($\tau=250$ fs).  
These yield $(\wpl \tau)/2\pi \simeq 8\;\text{THz}$, with values of
$E_0 \simeq 18\;\text{kV}/\text{cm}$ and $\omega/2\pi \simeq
4\;\text{THz}$ ($\hbar \omega \simeq 31\;\text{meV}$) required to
reach the plasma instability region. These values are beyond what is
reported in the literature for wide miniband and highly doped or low
$\gamma$ superlattices\cite{schomburg99}. On the other hand, previous
theoretical studies using self-consistently calculated energy
dependent scattering rates have had $\wpl \tau$ well in excess of
$~10$ near miniband center\cite{cao00:chaos-qdot-miniband}.  Due to
field screening effects, very high $\wpl$ requires correspondingly
high pump frequencies. For this reason single miniband transport model
can become questionable, since it requires that $\hbar \omega \ll
\Delta_g$, where $\Delta_g$ is the gap between first and second
minibands.

To gain better understanding of the dynamics and physics of the
instability, we transform our equations into nearly Hamiltonian
variables, \ie\ into a form that differs from a Hamiltonian by a small
parameter.  We rewrite our equations in terms of new variables $q$,
$p$, and $\Delta$:
\begin{subequations}
  \label{eq:pendvars}
  \begin{gather}
    V = \frac{a \Delta}{2\hbar} \sin \frac{a p}{\hbar}, \quad 
    W = -\frac{\Delta}{2}\cos \frac{a p}{\hbar}, \\
    \Esc = -\frac{4 \pi}{\eps_0} eNq.
  \end{gather}
\end{subequations}
Physically, $p$ is the mean value of the electron ensemble
(quasi)momentum while $\Delta$ describes an effective instantaneous
miniband width, reduced from $\Delta_0$ due to the spreading of the
electron distribution in momentum space. More precisely, the fraction
$\Delta/\Delta_0 = 1 - \cvar$, where $\cvar$ is the variance of the
momentum distribution (using a definition appropriate for periodic
distribution functions). By the virtue of being the opposite of the
variance, $\Delta/\Delta_0$ describes the coherence of the ensemble
motion, \ie\ the concentration (bunching) of the electrons to the
vicinity of the mean momentum $p$. The interpretation of these
variables is rigorously justified in Appendix~\ref{app:meanmom}.
 
Substituting the above to Eqs.~(\ref{eq:be}) we obtain
\begin{subequations}
  \label{eq:he}
  \begin{eqnarray}
    \dot \Delta &=& -\gamma \left[ \Delta + 2 \Weq \cos \frac{a p}{\hbar}\right], \label{eq:he-Delta} \\
    \dot p &=& e \Eext - \frac{4 \pi e^2 N}{\eps_0} q + 2 \gamma \frac{\hbar \Weq}{a \Delta} \sin \frac{a p}{\hbar}, \label{eq:he-p} \\
    \dot q &=& \frac{a \Delta}{2\hbar} \sin \frac{a p}{\hbar}. \label{eq:he-q}
  \end{eqnarray}
\end{subequations}
The new governing equations can be thought of as effective
semiclassical equations of motion for an electron ensemble. A notable
difference is that $\Delta$ is here a dynamic variable rather than a
constant, and that the self-consistent field, acting essentially as a
harmonic returning force, is present via the variable $q$.

The Hamiltonian limit, $\gamma \to 0$, of Eq.~(\ref{eq:he}) has a
well-studied classical analog. Defining the phase $\theta = a p /
\hbar$ one sees that $\theta$ follows the equation of a driven
pendulum:
\begin{equation}
  \ddot \theta + \wpl^2 \frac{\Delta}{\Delta_0} \sin \theta = 
  -\frac{e a E_0}{\hbar} \omega \sin \omega t.
  \label{eq:pend}
\end{equation}
Previous works have also considered the pendulum limit of the
superlattice balance equations\cite{alekseev02a, isohatala05:sbpend}.
These have included damping via a term $-\gamma \dot \theta$ on the
right-hand side to model the effect of dissipation, while supposing
$\Delta = \text{constant}$.  However, such a system cannot capture all
the dynamics observed in the third order differential
equations~(\ref{eq:he}). This is because for the damped second order
pendulum all phase space areas contract along the flow, making
Hopf-bifurcation impossible. This implies that the interaction
between the high- and low frequency parts, $(q, p)$, and $\Delta$,
respectively, generates the quasiperiodic oscillations.
\newcommand{\Ecrit}{E_\text{crit}}

Using Eqs.~(\ref{eq:he}) we have computed analytically approximate
conditions for the appearance of the Hopf-bifurcation in the weak
drive limit. We defer the details of the analysis to
Appendix~\ref{app:math}. Our findings show that the Hopf-bifurcation
occurs for $E_0 > \Ecrit$, where $\Ecrit \to 0$ as $\gamma \to 0$, and
for frequency
\begin{equation}
  \OmegaR + \order(\gamma^2) < \omega < \OmegaH + \order(\gamma^2),
  \label{eq:plina}
\end{equation}
where $\OmegaR$ is the amplitude dependent natural oscillation
frequency of the system, $\OmegaH > \OmegaR$ is an upper limiting
frequency for the Hopf-bifurcation, and $\order(\gamma^2)$ represent
terms proportional $\gamma^2$. Both $\OmegaR$ and $\OmegaH$ tend to
$\wpl$ as $E_0 \to 0$. The plasma instability is in other words first
triggered at a small positive detuning from the resonance $\omega =
\OmegaR$, and exists for a modest range of parameters extending to
higher frequencies from resonance, and can be brought about by even a
small $E_0$ provided that $\wpl \tau$ is large enough. The frequency
of the modulation of $\Delta$ and $(p,q)$ oscillation, denoted
$\Omega$, is small, vanishing at the resonance $\omega = \OmegaR$ and
increasing as parameters varied away from it.

\begin{figure*}
  \begin{center}
    \includegraphics[scale=0.95]{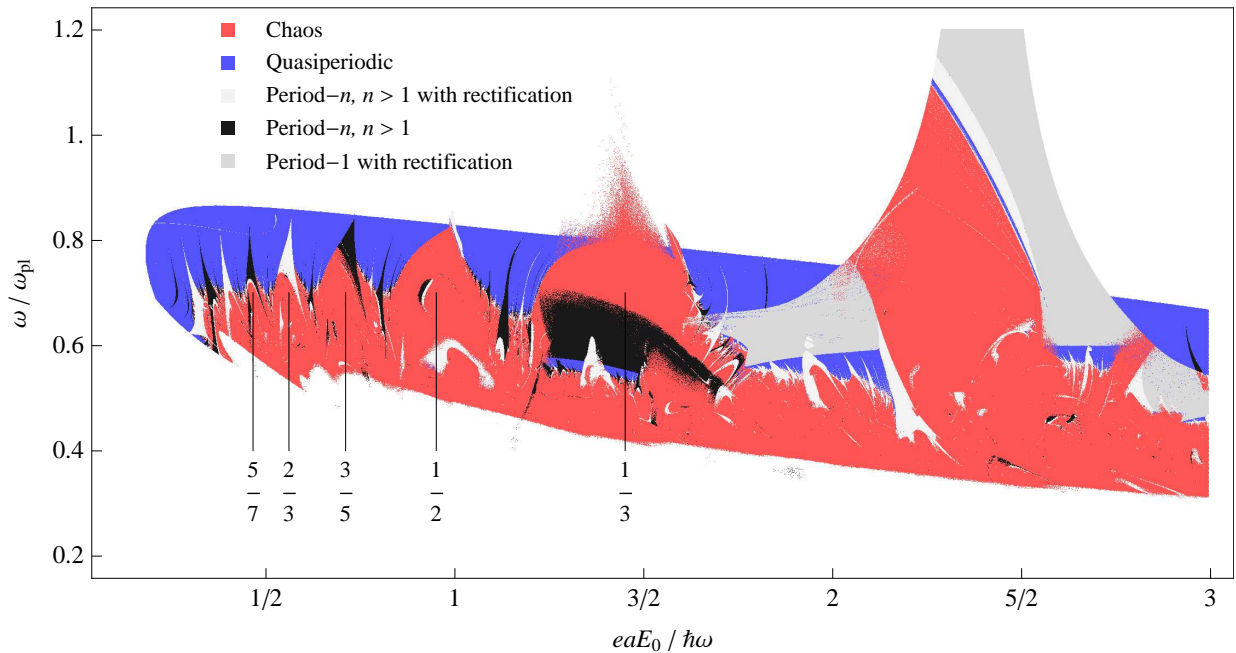}
  \end{center}
  \caption{\label{fig:mapwpl20} [Color online] Regions of different
    types of dynamics, $\wpl \tau = 20$. Quasiperiodic motion [blue],
    frequency-locking [light gray and black], or chaos [red] appears
    as a result of the plasma instability. For tongues with locking
    ratios $p/q$, indicated by the labels for select resonances, such
    that either $p$ or $q$ is even [light gray regions] a spontaneous
    \dc-voltage is present. Transition to chaos is associated with
    overlapping of the resonance tongues. The region of period-1
    solutions with rectification near $eaE_0 \sim 2.4 \hbar \omega$
    extending to high-frequencies outside the shown range is born out
    of the first \ndc-instability region.
  }
\end{figure*}

Turning to physics point of view, the analytic results show that the
most significant contributing factor to the appearance of the
instability is the coupling between the slow degree of freedom
$\Delta$ to the resonance frequency $\OmegaR$, and in turn, the down
mixing of the fast oscillations to the motion of $\Delta$. This is in
direct analogy to what is found for coupled high- and low-frequency
oscillators\cite{shygimaga98}, where a very feedback coupling between
slow and fast parts was found to lead to a Hopf-bifurcation followed
by transition to chaotic motion. Physically, $\Delta$ affects the
resonance frequency via the coherence $\Delta/\Delta_0$. This is
easily seen by noting that the resonance frequency is proportional to
the frequency of free linear oscillations, $\wpl
\sqrt{\Delta/\Delta_0}$ [cf. Eq.~(\ref{eq:pend})], essentially a
coherence-dressed plasma frequency. Thus, the slow motion of $\Delta$
couples to the fast oscillations by modulating the resonance
frequency, and thereby affecting a change in the amplitude of the $p,
q$ oscillations, with strongest response occuring near $\omega \sim
\OmegaR$.

The reverse coupling, \ie\ down mixing of fast to slow motion, is due
to scattering induced loss of coherence, described by
Eq.~(\ref{eq:he-Delta}). The rate of decoherence increses as the mass
of the electron distribution is offset from the band bottom, since
scattering events in the present model return electrons to the thermal
distribution that is centered to $p = 0$.  Given the change in
$\Delta$ is essentially adiabatic at the present range of parameters,
the decoherence per drive cycle is then dependent on the amplitude of
$p, q$ oscillations, with large amplitudes contributing a larger loss
of coherence. This way a slow modulation of the fast oscillation
amplitude is converted back to a variation in the coherence.

\section{Synchronization, rectification, and chaos}
The Hopf-bifurcation introduces a new frequency into the system by
essentially generating spontaneous oscillations that appear as a
parametric self-excitation, as can be seen from Eq.~(\ref{eq:pend})
and noting that $\Delta/\Delta_0$ is now a slowly oscillating
function. It is expected, then, that this generates new dynamics, and
in this section we discuss the various limiting dynamics that we have
found in numerical simulations of Eqs.~(\ref{eq:be}).

We have observed three different types of motion as a trajectory
evolves away from the unstable focus: quasiperiodic oscillations,
frequency locking, and chaos. In Fig.~\ref{fig:mapwpl20} we have
plotted the type of attractor corresponding to fixed initial
conditions of $V(0) = 0, W(0) = 0, \Esc(0) = 0$.  In this example,
trajectories ejected from the unstable focus-saddle typically converge
to quasiperiodic oscillations, making the attractor a torus. This
introduces a new frequency into the system that is incommensurate to
the drive frequency $\omega$. We denote this frequency by $\omega_2$,
and compute it by finding the strongest peak in Fourier transforms of
$\Esc$ that is not an integer multiple of $\omega$. The frequency
$\omega_2$ is related to the slow modulation frequency of the previous
section by $\omega_2 \simeq \omega \pm \Omega$. This relation is exact
only for small plasma oscillations, and typically it is the minus sign
that applies (frequency approximately $\omega + \Omega$ is present in
the frequency spectrum, but is much weaker).

Of particular interest is the application of the quasiperiodic
oscillations to detection of THz-radiation. To this end, spontaneous
\dc\ voltages such as those appearing at the \ndc-instability would be
desirable. However, the Fourier transforms of the net electric field
$E$ on the tori show no zero harmonic component. The Hopf-bifurcation
is super-critical and so the amplitude of the plasma oscillations is
small near the bifurcations, but grows rapidly away from the critical
curve. The frequency $\omega_2$ varies between $\wpl$ and $0$, with
the frequency decreasing as the \ndc-regions are approached.

\begin{figure}
  \begin{center}
    \includegraphics[scale=0.95]{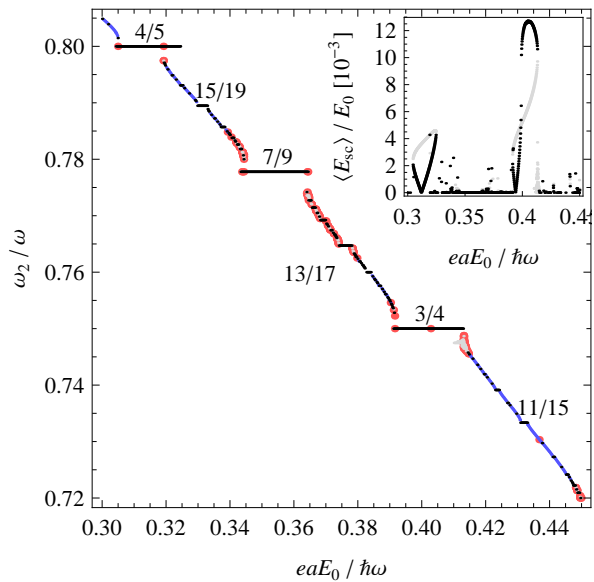}
  \end{center}
  \caption{\label{fig:dsc} Main figure: Frequency of the plasma
    oscillations as a function of applied field amplitude $E_0$, $\wpl
    \tau = 20$, $\omega \tau = 14.1$. Black and gray dots correspond
    to stable and unstable phase-locked solutions ($\omega_2/\omega$
    rational), respectively, while blue points mark irrational
    frequency ratios. Chaos [red] is seen to occur where the curve
    becomes dense with plateaus, e.g. near and between 7:9 and 13:17
    resonances. Inset: Generated spontaneous \dc-field, $\langle \Esc
    \rangle$ over the same range of $E_0$. Strong resonances at
    $\omega_2 = 3/4 \cdot \omega$ and $\omega_2 = 4/5 \cdot \omega$
    are clearly visible as relatively large peaks in the \dc\ bias.  }
\end{figure}

For decreasing $\omega$ the plasma oscillation amplitude increases.
This increases the coupling of $\omega$ and $\omega_2$ modes resulting
in synchronization of the plasma oscillations to the drive. When in
synchrony, the relationship
\begin{equation}
  p \omega = q \omega_2,
\end{equation}
where $p$ and $q$ are integers, holds for a finite range of
parameters.  For $\wpl \tau \gtrsim 10$, the frequency locking regions
appear as Arnol'd tongues in and around the quasiperiodic
regime\footnote{Numerical data of Ref.~\onlinecite{shygimaga98} shows
  traces of regions of synchronization. This phenomenon was not
  investigated, however, possibly because the numerical methods were
  not able to detect frequency locking.}. On the tongues the period of
the motion starts as $q$ times the period of the pump field $T$, and
increases via period doubling bifurcations as chaotic regime is
approached.  A more detailed picture is provided in
Fig.~\ref{fig:dsc}, where we have plotted $\omega_2 / \omega$ over a
range crossing multiple Arnol'd tongues. The curve bears the shape of
a classic Devil's staircase where the plateaus indicate frequency
locking.  Significantly, if either $q$ or $p$ is even, or if a period
doubling has occurred, we observe a spontaneous generation of a weak,
unquantized \dc\ bias. In the inset of Fig.~\ref{fig:dsc} we have
plotted the value of \dc-electric field across the superlattice,
$\langle E \rangle = \langle \Esc \rangle$. Non-zero values in the
\dc\ component can be seen at even valued resonances, however, no
quantization is present. The fact that no \ndc\ is observed, implies
that synchronization can be used in the detection of intense
THz-radiation. We note that tongues with locking ratios such that $p,
q \gg 1$ are seen only emerging inside the region of plasma
instability. On the other hand, for locking ratios 1:2 and 1:3 the
corresponding tongue survives outside the plasma instability
regime. Although the initial conditions used in
Fig.~\ref{fig:mapwpl20} prefer the period-1 solutions, these
synchronization regions persist to higher $\omega$ than the
quasiperiodic oscillations. This means that the 1:2 resonance can in
principle be used for THz detection via \ac\ to \dc\ bias conversion
in a wider range of parameters.

The resonance tongues form a border between quasiperiodic and chaotic
oscillations. For decreasing $\omega$, the mode-locked plateaus widen
and more of them appear, viz. Fig.~\ref{fig:mapwpl20}. Accordingly,
the staircase starts to show distinct fractal character as steps form
on shorter and shorter length scales. Transition to chaos is then
observed as nearby plateaus overlap. This is seen in
Fig.~\ref{fig:dsc} for example near 3:4 resonance and between 7:9 and
13:17 resonances.  Note that the resonance plateaus are still present,
but unstable and embedded in the chaotic attractor. Onset of chaos in
this system can therefore be seen as following the overlapping of
nearby resonances\cite{chirikov79}, implying the collision of the
heteroclinic cycles connecting the saddle points associated with the
$q$-periodic orbit.

The route to chaos as observed here is a ``universal'' type of route
to chaos also in dissipative, drive oscillatory systems
\cite{jensen84, *bohr84, knudsen91, feingold88}, and dissipative
systems with two competing frequencies. These include systems
described by the \ac+\dc\ driven and damped pendulum equation, such as
Stewart-McCumber \cite{stewrt, *mccumber} model of Josephson
junctions, and systems reducing to circle maps or similar dissipative
standard-like maps. A notable difference here is that the second
frequency is self-generated via the Hopf-bifurcation, and not \eg\ an
additional external drive term.

\section{Summary}
We have performed an extensive study of the dynamics of the plasma
instability in semiconductor superlattices under intense
THz-radiation. We found an analog of the instability occurring for
wide-miniband, highly doped superlattices that was studied in
Ref.~\onlinecite{romanov79, *romanov79-trans}. This instability is
different from that associated with negative differential
conductivity, and appears for parameters where electron transport is
expected to be stable against domains.

The instability leads to sustained oscillations that are quasiperiodic
at some $\omega_2$, frequency locked whereby $\omega_2$ is fixed to a
rational multiple of pump-field, $\omega_2 = (p/q) \omega$, or
chaotic.  Importantly from an applications point of view, the
synchronization of the plasma oscillations to the drive, $p \omega = q
\omega_2$, with an even $p$ or $q$, generates a spontaneous
\dc\ voltage that can be used in detection of THz radiation. In
particular, the $p = 1$, $q = 2$ resonance appears particularly
strong, and survives for a wide range of parameters.

Transition to chaos was found to occur as resonance tongues overlap,
showing that the plasma instability is a precursor to the chaotic
dynamics found in semiconductor superlattices in previous works.
Indeed, it appears that this instability underlies much of the complex
dynamics observed in highly-doped semiconductor superlattices.
Finally, we also elucidated the physical mechanism sustaining the
plasma oscillations. We found that scattering induced down-mixing of
high-frequency oscillations into a slow modulation of resonance
frequency of free miniband electrons is mainly responsible for the
appearance of the instability.

\newcommand{\Ep}{E_\text{probe}}
\newcommand{\Epa}{{\tilde E}}

\appendix
\section{\label{app:math}Details of mathematical analysis}
\subsection{Negative differential conductivity implies instability}
Consider a symmetric limit cycle and introduce an $E$-field probe of
the form $\Epa_\Omega \cos(\Omega t)$ where $\Omega$ is an arbitrary
probe frequency and $\Epa$ is the probe amplitude. $\Epa$ is assumed
to be slowly varying. The probe induces a change in the current
density of the form $j \to j + \pdin{j}{\Epa_\Omega}$. So being, we
can write the time-evolution of $\Epa$ from Eq.~(\ref{eq:be-Esc}) as
\begin{equation}
  {\dot \Epa}_\Omega \cos \Omega t - \Omega \Epa_\Omega \sin \Omega t 
  = 
  -\frac{4\pi}{\eps_0}\pd{j}{\Epa_\Omega}\Epa_\Omega.
\end{equation}
Multiplying by $\cos \Omega t$ and averaging over $T$ one obtains Eq.
\begin{equation}
  {\dot \Epa}_\Omega = -\frac{4\pi}{\eps_0}\Re\{\sigma_\Omega\} \Epa_\Omega,
\end{equation}
where $\sigma_\Omega$ is the differential conductivity at frequency
$\Omega$, $\sigma_\Omega = 2 \pdin{\langle j \exp(\I \Omega t)
  \rangle}{\Epa_\Omega}$, $\Omega > 0$, and $\sigma_0 = \pdin{\langle
  j \rangle}{\Epa_0}$. The damping rate of the probe field is then
$(4\pi/\eps_0)\Re\{\sigma_\Omega\}$, and thus, if $\sigma_\Omega < 0$,
the probe is unstable, and if $\Omega > 0$, the corresponding
limit-cycle is clearly a unstable focus-saddle. For $\sigma_0 < 0$ the
local divergence is exponential and not oscillatory, making the
unstable point a saddle.

\subsection{Plasma instability}
Here we present the details of the analytic formulas for the
appearance of the plasma instability. For simplicity, we consider here
the limit $e E_0 \ll \hbar \omega$. We approximate $(p, q)$ with a
single $\omega$ harmonic with slowly varying amplitude and phase $R,
\Phi$,
\begin{gather}
  p = \frac{\hbar R}{a} \sin (\omega t + \Phi), \quad
  q = \frac{\hbar \omega \eps_0 R}{4\pi e^2 N} \cos (\omega t + \Phi).
\end{gather}
We substitute these into Eqs.~(\ref{eq:he}), solve for the derivatives
of the new variables, and finally, apply the averaging
method\cite{verhulst05} to get the following equations for $\Delta,
R$, and $\Phi$:
\begin{eqnarray}
  \dot \Delta &=& -\gamma(\Delta - \Delta^*(R)), \\
  \dot R &=& -\Gamma R + \frac{aeE_0}{2\hbar} \sin \Phi, \\
  \dot \Phi &=& -\frac{1}{2}\left(\omega - \frac{\OmegaR}{\omega} \right)
  + \frac{a e E_0}{2\hbar R}\cos \Phi,
  \label{eq:slow}
\end{eqnarray}
where $J_k$ is the $k$th order Bessel function of the first kind,
$\Delta^*(R) = 2 |\Weq| J_0(R)$, $\Gamma(\Delta,R) = 2\gamma |\Weq|
J_1(R)/R\Delta$, and $\OmegaR$ is the free, unforced, undamped
oscillation frequency,
\begin{equation}
  \OmegaR(R, \Delta)^2 = \wpl^2 \frac{\Delta}{\Delta_0}\frac{2J_1(R)}{R}.
  \label{eq:OmegaR}
\end{equation}
Approximate symmetric limit cycles are given by steady-state values of
$R, \Phi$, and $\Delta$. The phase $\Phi$ can be eliminated from the
equations, while $\Delta = \Delta^*(R)$ and $R$ satisfies the implicit
equation
\begin{gather}
  R^2 
  =
  \left[ \frac{e a E_0}{\hbar} \right]^2
  \frac{1}{(2\Gamma)^2 + \left( \omega - \frac{\OmegaR^2}{\omega} \right)^2}.
  \label{eq:reso}
\end{gather}
Dynamics near the equilibrium are determined by the Jacobian of the
left-hand side of Eqs.~(\ref{eq:slow}). Frequency of small plasma
oscillations is found up to first order in $\gamma$ to be
\begin{equation}
  \Omega = 
  \sqrt{
    \left( \frac{\omega}{2} - \frac{\OmegaR^2}{2\omega} \right)
    \left( \frac{\omega}{2} - \frac{\OmegaS^2}{2\omega} \right)
  } + \order(\gamma^2),
\end{equation}
where the second frequency $\OmegaS$ arises from $R$ dependence of the
$\OmegaR$,
\begin{gather}
  \OmegaS^2
  = \wplp^2 J_0(R)\left[ J_0(R) - J_2(R) \right],
\end{gather}
where $\wplp^2 = 2|\Weq|\wpl^2/\Delta_0$.  Frequencies $\OmegaR$ and
$\OmegaS$ satisfy $\OmegaS < \OmegaR$ for all $\omega$ and $R$, thus,
the frequency $\Omega$ is real for $\omega < \OmegaS$ and $\omega >
\OmegaR$.

Instability regions are given by applying the standard Routh-Hurwitz
stability criterion to the characteristic polynomial of the Jacobian:
\begin{subequations}
  \label{eq:rh}
  \begin{gather}
    \frac{1}{4}\left(\omega - \frac{\OmegaR}{\omega} \right)
    \left(\omega - \frac{\OmegaH}{\omega} \right)
    + (2\gamma - \Gamma) (\Xi + \Gamma)
    < 0, \label{eq:rh-plasma} \\
    \frac{1}{4}\left(\omega - \frac{\OmegaR}{\omega} \right)
    \left(\omega - \frac{\OmegaM}{\omega} \right) 
    + \Gamma (\Xi - \Gamma)
    < 0, \label{eq:rh-multi}
  \end{gather}
\end{subequations}
where $\OmegaH > \OmegaR$ and $\OmegaM < \OmegaR$ are $\gamma \to 0$
upper Hopf-bifurcation limit and lower nonlinear resonance limit,
respectively,
\begin{equation}
  \OmegaHM^2 = \OmegaS^2 \pm 2\wplp^2 J_1(R)^2,
  \label{eq:OmegaHM}
\end{equation}
and
\begin{gather}
  \Xi = \gamma \left[1 - \frac{J_1(R)}{RJ_0(R)^2}(J_0(R) + R J_1(R)) \right].
\end{gather}
First of the inequalities~\ref{eq:rh} is true in the presence of the
plasma instability, while the second holds for the instability arising
from the nonlinear resonance. In short, Eqs.~(\ref{eq:rh}) describe
two instability bands, one below resonance (arising from
multistability) and second above resonance (plasma instability).
Noting that the second term in both inequalities is
$\order(\gamma^2)$, Eq.~(\ref{eq:plina}) follows.

Finally, the second term in Eq.~(\ref{eq:OmegaHM}), is critical for
the plasma instability. If it were too small, $\OmegaH$ would fall
below $\OmegaR$, and the instability would not appear. Looking
in an earlier stage of the derivation, this term equals
\begin{equation}
  \alpha = 2 R \OmegaR \pd{\Delta^*}{R} \pd{\OmegaR}{\Delta}.
\end{equation}
This crucially depends on the strength of the coupling of the
amplitude of the fast oscillations to $\Delta$ ($\pdin{\Delta^*}{R}$
coefficient), and the coupling $\Delta$ to the free oscillations
frequency ($\pdin{\OmegaR}{\Delta}$ coefficient). This leads us to
conclude that the down-mixing of the fast motion in the equation for
$\Delta$ and the subsequent modulation of the free oscillation
frequency is the main physical mechanism generating and sustaining the
plasma oscillations.

\section{\label{app:meanmom}$p$ is the ensemble mean quasimomentum, $\Delta/\Delta_0$ coherence}
Here we justify our claim that $p$ as introduced in
Eqs.~(\ref{eq:pendvars}) is the center of mass of the electron
distribution in (quasi)momentum space, while $\Delta/\Delta_0$
describes the coherence of Bloch oscillations, or the concentration
(bunching) of the electrons to vicinity of the mean momentum. We will
use the short-hand notations $\theta = ap/\hbar$ and $A =
\Delta/\Delta_0$.

Let $f(\pz, t)$ be the electron momentum distribution function as
given by the Boltzmann transport equation, where $\pz$ is the momentum
along the direction of the current, and let overline denote averaging
against $f(\pz, t)$, $\overline{(\cdot)} = (a/\pi
\hbar)\int_{-\pi\hbar/a}^{\pi\hbar/a}(\cdot)f(\pz,t)\;\D \pz$.  Due to
the miniband structure, $f(\pz, t)$ is a periodic function of $\pz$,
and for conveniance we express $\pz$ by an angle $\phi$, $\phi =
a\pz/\hbar$, $-\pi < \phi \leq \pi$.
We first note that the naïve expectation value of the angle
$\overline{\phi}$, cannot correctly represent the center of mass
position, and consequently $\overline{p}_z$ is not a suitable choice
as the mean momentum. This becomes clear by considering $f(\phi,t)$
that is condensed to a narrow, symmetric peak at $\phi = \pi$. In such
a case, the expectation value $\overline{\phi} = 0$, in complete
opposition to the ``true'' value of $\pi$.

A more appropriate definition for the mean angle, which we will denote
by $\cang$, is given via the first trigonometric moment $m_1$ of the
distribution\cite{mardia}:
\begin{equation}
  \cang = \arg m_1, \quad m_1 = \overline{\exp(\I \phi)},
  \label{eq:cang}
\end{equation}
where $\arg z$ gives the complex argument of $z$. Furthermore, in
analogy to the use of variance of distributions on the real line, the
spread of a circular distribution can be characterized by its circular
variance $\cvar$,
\begin{equation}
  \cvar = 1 - |m_1|.
  \label{eq:cvar}
\end{equation}
Note that $0 \leq \cvar \leq 1$. Zero $\cvar$ is equivalent to
the entire distribution being concentrated to the mean angle $\cang$,
while unity $\cvar$ implies that a well-defined mean angle does not
exist.

We show next that $\cang$ and $\cvar$ coincide with $\theta$ and $1 -
A$. Returning to the definition of $V$ and $W$,
\begin{gather}
    V = V_\text{max} \overline{\sin (a \pz / \hbar)}, \quad
    W = -\frac{\Delta_0}{2} \overline{\cos (a \pz / \hbar)}.
    \label{eq:vwdef}
\end{gather}
From Eqs.~(\ref{eq:pendvars}) and~(\ref{eq:vwdef}) one finds that
\begin{gather}
  A \sin \theta = \overline{\sin (a\pz/\hbar)}, \quad
  A \cos \theta = \overline{\cos (a\pz/\hbar)},
\end{gather}
or equivalently $A\exp(\I \theta) = \overline{\exp(\I \phi)}$.  It
then follows directly from the definitions, Eqs.~(\ref{eq:cang})
and~(\ref{eq:cvar}) that $\theta = \cang$ and $\cvar = 1 - A$.
Scaling angles back to momentum units, we have $p^* = \hbar \cang / a
= \hbar \theta / a = p$, proving our assertion that $p$ is the center
of mass momentum.

\bibliography{bibliography}
\end{document}